\documentclass[aps,prl,twocolumn,showpacs,a4paper]{revtex4-1}
\pdfoutput=1

\usepackage{bm}
\usepackage{mathtools}
\usepackage{amssymb}
\usepackage{placeins}
\usepackage[left]{lineno}

\usepackage{geometry}
\renewcommand{\vec}[1]{\ensuremath{\bm{#1}}}

\newcommand\Li{\ensuremath{{^6}\mathrm{Li}}}

\begin{document}

\title{Experimental realization of a long-range antiferromagnet in the Hubbard model with ultracold atoms}

\author{A. Mazurenko}
\author{C. S. Chiu} 
\author{G. Ji} 
\author{M. F. Parsons}
\author{M. Kan\'asz-Nagy}
\author{R. Schmidt}
\author{F. Grusdt}
\author{E. Demler}
\author{D. Greif}
\author{M. Greiner}
\email{greiner@physics.harvard.edu}
\affiliation{Department of Physics, Harvard University, Cambridge, Massachusetts 02138, USA}


\maketitle

\textbf{Many exotic phenomena in strongly correlated electron systems emerge from the interplay between spin and motional degrees of freedom \cite{Auerbach1997, Sachdev2011}. 
For example, doping an antiferromagnet gives rise to interesting phases including pseudogap states and high-temperature superconductors \cite{Lee2006}. 
A promising route towards achieving a complete understanding of these materials begins with analytic and computational analysis of simplified models. 
Quantum simulation has recently emerged as a complementary approach towards understanding these models \cite{Kim2010, Struck2011, Simon2011, Drewes2016, Murmann2015}. 
Ultracold fermions in optical lattices offer the potential to answer open questions on the low-temperature regime of the doped Hubbard model \cite{Hofstetter2002, Bloch2008, Esslinger2010}, which is thought to capture essential aspects of the cuprate superconductor phase diagram but is numerically intractable in that parameter regime. 
Already, Mott-insulating phases and short-range antiferromagnetic correlations have been observed, but temperatures were too high to create an antiferromagnet \cite{Jordens2008, Schneider2008, Greif2013b, Hart2014}. 
A new perspective is afforded by quantum gas microscopy \cite{Bakr2009, Sherson2010, Haller2015, Cheuk2015, Parsons2015, Edge2015a, Omran2015, Greif2016, Cheuk2016b, Parsons2016, Boll2016, Cheuk2016a, Brown2016}, which allows readout of magnetic correlations at the site-resolved level \cite{Parsons2016, Boll2016, Cheuk2016a, Brown2016}. 
Here we report the realization of an antiferromagnet in a repulsively interacting Fermi gas on a 2D square lattice of approximately 80 sites. 
Using site-resolved imaging, we detect (finite-size) antiferromagnetic long-range order (LRO) through the development of a peak in the spin structure factor and the divergence of the correlation length that reaches the size of the system.
At our lowest temperature of $\mathbf{T/t=0.25(2)}$ we find strong order across the entire sample, where the staggered magnetization approaches the ground-state value. Our experimental platform enables doping away from half filling, where pseudogap states and stripe ordering are expected, but theoretical methods become numerically intractable. In this regime we find that the antiferromagnetic LRO persists to hole dopings of about $\mathbf{15\%}$, providing a guideline for computational methods. Our results demonstrate that quantum gas microscopy of ultracold fermions in optical lattices can now address open questions on the low-temperature Hubbard model.
}

The Hubbard Hamiltonian is a fundamental model for spinful lattice electrons describing a competition between kinetic energy $t$ and interaction energy $U$ \cite{Hubbard1963}.
In the limiting case of half-filling (average one particle per site) and dominant interactions ($U/t \gg 1$) the Hubbard model maps to the Heisenberg model \cite{Auerbach1997}. 
There, the exchange energy $J=4t^2/U$ can give rise to antiferromagnetically ordered states at low temperatures \cite{Manousakis1991}.
This order persists for all finite $U/t$, where charge fluctuations reduce the ordering strength \cite{Hirsch1985}.
Away from half-filling, the coupling between motional and spin degrees of freedom is expected to give rise to a rich many-body phase diagram (see Fig.~1a), which is challenging to understand theoretically due to the fermion sign problem \cite{Scalapino2006}.
Even so, in the thermodynamic limit commensurate long-range order (LRO) has been conjectured to transition to incommensurate LRO infinitesimally far from half-filling, whereas for finite-size systems commensurate order is expected to extend to non-zero doping \cite{Hirsch1985, Schulz1990}.

\begin{figure}[t!]
\begin{center}
	\includegraphics{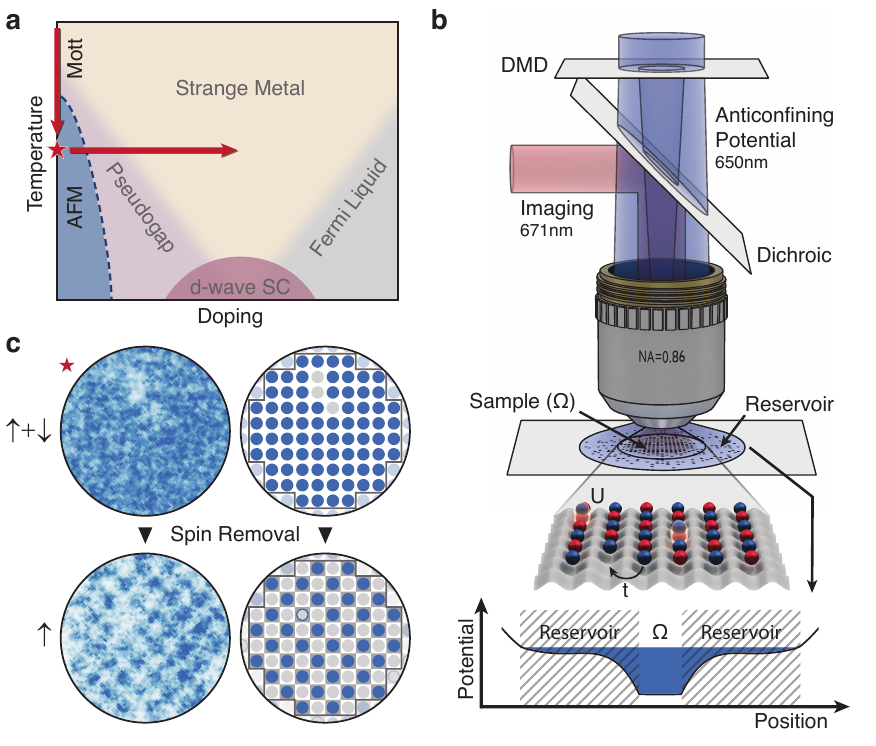}
	\label{fig:intro}
\end{center}
	\caption{\textbf{Probing antiferromagnetism in the Hubbard model with a quantum gas microscope.} 
		\textbf{a,} Schematic view of the 2D Hubbard phase diagram, including predicted phases. This work explores the trajectories traced by the red arrows for a $U/t=7.2(2)$ Hubbard model. The strongest antiferromagnetic order is observed at the starred point.  
		\textbf{b,} Experimental setup. We trap $\Li$ atoms in a 2D square optical lattice. We use the combined potential of the optical lattice and a DMD to trap the atoms in a central sample $\Omega$ of homogeneous density, surrounded by a dilute reservoir.
		\textbf{c,} Exemplary raw and processed images of the atomic distribution of single experimental realizations, with both spins components present (upper) and one spin component removed (lower). The observed chequerboard pattern in the spin-removed images indicates the presence of an antiferromagnet.}
\vspace{1cm}
\end{figure}

The strength of global antiferromagnetic order in spin systems on bipartite lattices is quantified by the staggered magnetization $m=|\vec{m}|$. 
The component along the $z$ spin direction is
\begin{equation}
\label{eq:stag-magn}
m^z = \sqrt{\langle \left(\hat{m^z} \right)^2 \rangle} = \sqrt{\left \langle \left(\frac{1}{N}\sum_i (-1)^i \frac{1}{S}\hat{S}^z_i \right)^2 \right \rangle},
\end{equation}
with $m^2=(m^x)^2+(m^y)^2+(m^z)^2=3(m^z)^2$ for SU(2) symmetry \cite{Hirsch1985}. 
Here $\hat{S}^z_i$ is a spin-$S$ operator on lattice site $i$ and $N$ denotes the number of lattice sites. 
While the ground state of the classical Heisenberg model on a square lattice is a perfectly ordered N\'{e}el state with $m=1$, the situation is much more interesting for the quantum case: quantum superposition states such as local singlet pairs reduce the energy of the many-body state as compared to the classical case. These quantum corrections decrease the staggered magnetization in the ground state to $m=0.61$ (i.e. $m^z=0.35$) for the $S=1/2$ Heisenberg model \cite{Sandvik1997}. 
In 2D, LRO disappears ($m=0$) in the thermodynamic limit for finite temperatures, as stated by the Mermin-Wagner-Hohenberg theorem \cite{Mermin1966, Hohenberg1967}. 
There the spin correlations decay exponentially over a correlation length $\xi$, which grows exponentially with inverse temperature ($k_B=1$)
\begin{equation}
\label{eq:corr-length}
\xi(T) = C_{\xi} \exp\left(\frac{2\pi\rho_s}{T}\right),
\end{equation}
where $\rho_s$ is the spin stiffness and $C_{\xi}$ is a constant \cite{Manousakis1991}. However, for the finite-size system investigated in this work, a crossover to antiferromagnetic long-range order does occur at a non-zero temperature, where $\xi$ becomes comparable to the system size and $m^z$ becomes of order unity. 

Two aspects were critical in realizing antiferromagnetic LRO in our experiment: first, reaching sufficiently low temperatures and second, creating a well-defined region of uniform density within the atomic cloud where the LRO state can form. 
We address both challenges simultaneously by exploiting the high-resolution microscope at the heart of the experiment, which enables in situ, high-fidelity, and site-resolved measurements of the lattice occupation.
We use a digital micromirror device (DMD) as a spatial light modulator in the image plane of the microscope to control the atomic potential landscape at a single-site level \cite{Liang2010}.
We engineer the potential to split the system into two subsystems: a central disk-shaped region $\Omega$ containing $>75$ sites, surrounded by a large reservoir at much lower density, see Fig.~1b (and Extended Data Fig.~1). 
Partitioning the system enhances the inherent entropy redistribution in the trap by shifting a higher fraction of the total entropy to the reservoir \cite{Ho2009a}.
Additionally, the potential within $\Omega$ is shaped to cancel the underlying harmonic potential, ensuring a highly uniform and tunable filling, see Extended Data Fig.~2 (Methods). 

A balanced mixture of the two lowest hyperfine states of $\Li$ with repulsive contact interactions is adiabatically loaded into an isotropic, square, $7.4(1)\,E_R$ lattice with spacing $a=569\,\mathrm{nm}$, where $E_R/h=25.6\,\mathrm{kHz}$. The lattice is combined with a DMD-engineered potential at the focus of the microscope.
The system is well described by the Hubbard model with $t/h=0.90(2)\,\mathrm{kHz}$ and $U/h=6.50(3)\,\mathrm{kHz}$ where $h$ is the Planck constant, leading to $U/t=7.2(2)$.  
Similar to previous work, our detection method is based on selective spin removal followed by site-resolved imaging of the remaining atomic distribution \cite{Parsons2016}, see Fig.~1c. 
Averaging over many independent experimental realizations, we determine the spin correlator along the $z$-direction
\begin{equation}
C_{\vec{d}} = \frac{1}{\mathcal{N}_{\vec{d}}} \frac{1}{S^2} \sum_{\substack{\vec{r},\vec{s} \in \Omega\\\vec{d}=\vec{r}-\vec{s}}} \langle \hat{S}_{\vec{r}}^z \hat{S}_{\vec{s}}^z \rangle - \langle \hat{S}_{\vec{r}}^z \rangle \langle \hat{S}_{\vec{s}}^z \rangle
\end{equation}
where the normalization $\mathcal{N}_{\vec{d}}$ is the number of different two-point correlators at displacement $\vec{d}$ within $\Omega$. 
The correlator compares the number of parallel and anti-parallel spin orientations on two sites separated by $\vec{d}$, i.e. is positive (negative) if parallel (anti-parallel) spin orientations are preferred.
Figure~2a shows $C_{\vec{d}}$ for different temperatures. 
For the lowest temperature we find spin correlations alternating in sign even up to the largest distance of $d=|\vec{d}|=10$ across the entire disk, as expected for an antiferromagnetic LRO state. 
We determine the temperature of each sample by comparing the measured nearest-neighbour correlator $C_{1}$ to quantum Monte Carlo predictions at half-filling, which gives $T/t=0.25(2)$ for the lowest temperature (Methods). 

\begin{figure*}[t]
\centering
	\includegraphics[width=\linewidth]{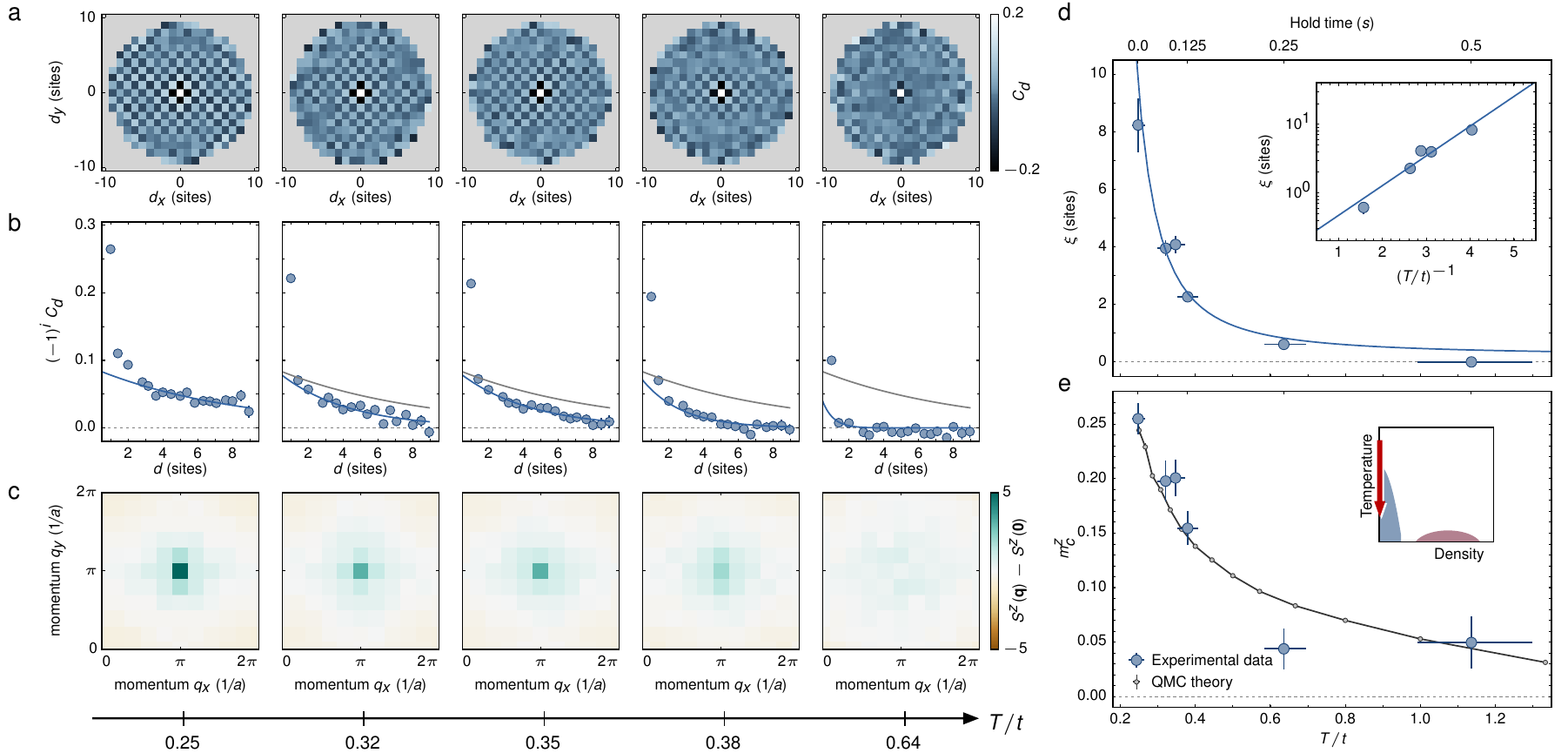}
	\label{fig:heating}
	\caption{\textbf{Observing antiferromagnetic long-range order.} 
		\textbf{a,} The spin correlator $C_{\vec{d}}$ is plotted for different displacements $\vec{d}$ ranging across the entire sample for five temperatures $T/t$. We record $>\!200$ images for each temperature (Methods). Correlations extend across the entire sample for the coldest temperatures, whereas for the hottest temperature only nearest-neighbour correlations remain.
		\textbf{b,} The sign-corrected correlation function $(-1)^iC_d$ is obtained through an azimuthal average. The exponential fits to the data ($d>2$) are shown in blue, from which we determine the correlation length $\xi$, and the fit of the coldest sample is plotted in grey for comparison. 
		\textbf{c,} The measured spin structure factor obtained from Fourier transformations of single images. A peak at momentum $\vec{q}_{\mathrm{AFM}}=(\pi/a,\pi/a)$ signals the presence of an antiferromagnet. 
		\textbf{d,} The measured correlation length $\xi$, fitted to Eq. (2), diverges exponentially as a function of temperature, and is comparable to the system size for the lowest temperature. The inset is a semi-logarithmic plot of the same quantity versus inverse temperature. 
		\textbf{e,} The measured staggered magnetization $m_z$ increases drastically below temperatures $T/t\approx 0.4$. We find good agreement with quantum Monte Carlo calculations of the Hubbard model, shown in grey. Error bars are computed as in (Methods). }
\end{figure*}

As temperature increases, the strength of antiferromagnetic order disappears rapidly, until for $T/t=0.64(6)$ only nearest-neighbour spin correlations remain. 
For a quantitative analysis of the spin correlations we plot in Fig.~2b a binned azimuthal average of the sign-corrected spin correlator $(-1)^iC_{d}$ as a function of distance $d$ (Methods). 
For large distances $d>2$ the measured correlation functions exhibit an exponential scaling with distance, verified by fitting $N_0\exp(-d/\xi)$ to each dataset, with the correlation length $\xi$ and $N_0$ as free parameters ($N_0$ the same for all fits). 
For our 2D system quantum fluctuations lead to an increase in spin correlations at short distances $d\leq2$ above the exponential dependence, most prominently visible in the nearest-neighbour correlator \cite{Gorelik2012}. 
In Fig.~2d we show the experimentally determined correlation length as a function of temperature, which increases dramatically at temperatures around $T/t=0.4$. 
For the lowest temperature we find $\xi=8.3(9)$ sites, which is approximately equal to the system size of $10$ sites, as expected for LRO. 

The long-wavelength and low-temperature behaviour of our system is expected to be well described by the quantum non-linear $\sigma$ model \cite{Baeriswyl1995, Auerbach1997}, which contains three fundamental ground-state parameters: the sublattice magnetization $M$, the spin stiffness constant $\rho_s$, and the spin-wave velocity $c$.
The spin stiffness quantifies the rigidity of an ordered spin system upon a twist \cite{Chakravarty1989, Chubukov1994, Caffarel1994}, and has been calculated to be $\rho_s/t\approx0.13$ for $U/t=7$, slightly below its Heisenberg value \cite{Denteneer1993}. 
Since the temperatures and correlation lengths are independently determined in our experiment, we can directly obtain an experimental value of $\rho_s$ by fitting the dependence in Eq.~\eqref{eq:corr-length} to the data. 
The data shows excellent agreement with the predicted exponential scaling of $\xi$ with $T^{-1}$ from Eq.~\eqref{eq:corr-length}.
From the fit we determine $\rho_s/t=0.16(1)$, which is larger than the calculated value, possibly due to finite-size effects (Methods).

Antiferromagnetic LRO in solid state systems is typically detected by neutron scattering or magnetic x-ray scattering \cite{Platzman1970, Squires2012}.
These methods measure the spin structure factor, given by
\begin{equation}
\label{eq:structure}
S^z(\vec{q})= \frac{1}{N}\sum_{\vec{r}, \vec{s} \in \Omega}^{N}\frac{1}{S^2}\langle \hat{S}^z_{\vec{r}} \hat{S}^z_{\vec{s}}\rangle\exp(i\vec{q}\cdot(\vec{r}-\vec{s})). 
\end{equation}
along the $z$-direction.
In a square lattice, antiferromagnetic LRO manifests as a peak in the structure factor at $\vec{q}_{\mathrm{AFM}}=\left(\pi/a, \pi/a\right)$, whose amplitude is directly related to the staggered magnetization $m^z=\sqrt{S^z(\vec{q}_{\mathrm{AFM}})/N}$.
For cold atom systems the spin structure factor can be measured from noise correlations or Bragg scattering of light \cite{Hart2014}. 
The site-resolved detection in our experiment allows for a direct measurement of the spin structure factor, obtained from averaging the squared Fourier transformation of individual single-spin images (Methods). 
The same result is obtained when summing over all contributions of the spin correlation function, see Extended Data Fig.~3.

For the lowest temperature we observe a sharp peak in the structure factor at $\vec{q}=\vec{q}_{\mathrm{AFM}}$, which confirms the presence of antiferromagnetic LRO, see Fig.~2c. 
For increasing temperatures the amplitude of this peak decreases until it disappears for $T/t\gtrsim 0.64$, indicating the decay of LRO. At these elevated temperatures a broad peak with low amplitude remains, which originates from the remaining short-range spin correlations. 
We quantify the ordering strength of the antiferromagnetic LRO by the corrected staggered magnetization $m^z_c(T)$, which subtracts uncorrelated contributions and is equal to $m^z$ in the thermodynamic limit (Methods). 
While initially small at elevated temperatures, $m^z_c$ shows a drastic increase for lower temperatures, see Fig.~2d. 
We compare the measured temperature dependence to \textit{ab initio} quantum Monte Carlo calculations of the Hubbard model on a $10\times10$ site square lattice with periodic boundary conditions and no free parameters.  
We find agreement over the entire range of temperatures, with residual deviations possibly caused by the different spatial shape of $\Omega$. 
The largest measured value of  $m^z_c=0.25(1)$ is more than $50\%$ of the theoretically predicted zero-temperature value in the Heisenberg model for our system size, obtained from finite-size scaling \cite{Sandvik1997}. 

\begin{figure}[t]
\begin{center}
	\includegraphics{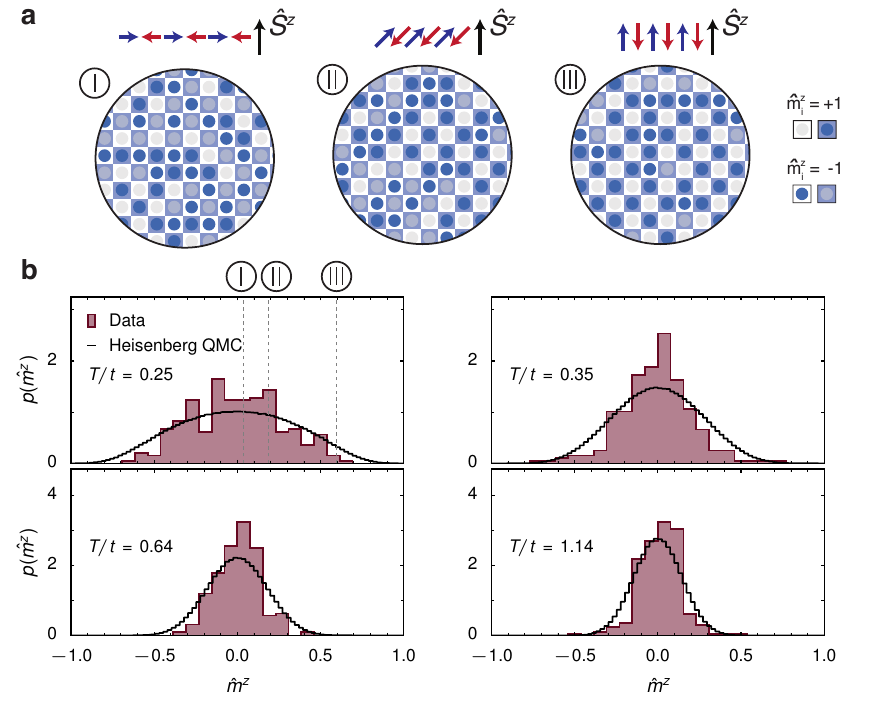}
\end{center}
	\caption{\textbf{Full-counting statistics of the staggered magnetization operator $\mathbf{\hat{m}_z}$.} 
		\textbf{a,} Selected images with one spin component removed (chequerboard overlaid to guide the eye) show a large variation in ordering strength at the coldest temperature. This variation is a consequence of the SU(2) symmetry of the underlying Hamiltonian, which leads to different orientations of the staggered ordering vector relative to the measurement axis $z$, as shown schematically by the spin-vectors. 
		\textbf{b,} Measured distributions of $\hat{m}_z$ are plotted at different temperatures. We find excellent agreement with quantum Monte Carlo simulations of the Heisenberg model with no free fitting parameters. 
		\label{fig:histogram}}
\vspace{1cm}
\end{figure}

The underlying Hubbard Hamiltonian that describes our system is SU(2) symmetric. 
In the absence of a symmetry-breaking field, the staggered spin-ordering vector $\hat{\vec{m}}=(\hat{m}^x,\hat{m}^y,\hat{m}^z)$ is expected to point in random directions on a sphere between different experimental realizations \cite{Auerbach1997}.
Consequently, individual measurements of the projection $\hat{m}^z$ are expected to show a large variation. 
This is directly observable in our experiment, as we can measure independent values of the staggered magnetization operator $\hat{m}^z$ from single experimental realizations, see Fig.~\ref{fig:histogram}a for a selection of images showing a large variation in the staggered ordering. 

The variation of the staggered ordering can be quantified from a histogram of all measured values of $\hat{m}^z$ across different experimental realizations, which corresponds to the full-counting statistics (FCS) of the operator $\hat{m}^z$. 
The FCS provide a powerful tool to characterize many-body systems beyond average values \cite{Hofferberth2008}, but so far has not been measured for the antiferromagnetic phase. 
Fig.~\ref{fig:histogram}b shows the measured histograms of the staggered magnetization along the $z$-direction for different temperatures at half-filling, obtained from over $250$ images each. 
All distributions are symmetric and peaked around zero with expectation values $\langle\hat{m}^z \rangle$ consistent with zero.
Additionally, we find the same results when measuring along a spin direction perpendicular to the $z$-axis via a $\pi/2$-pulse, see Extended Data Fig.~4.
Both observations are consistent with a randomly oriented ordering vector. 
The width of the distributions is characterized by the standard deviation $m^z$ defined in Eq.~\eqref{eq:stag-magn}.
At the highest temperature the distribution is consistent with the infinite temperature expectation, where the entire finite-size sample of $N$ sites is uncorrelated. 
There, a binomial distribution is predicted with a width $m^z(T\rightarrow \infty)=1/\sqrt{N}=0.1125$, which agrees with the experimentally measured value $m^z=0.12(2)$.
At lower temperatures the width of the distribution grows substantially and sensitively depends on temperature, but still remains peaked around zero. 
In the limit of vanishing temperatures and large system sizes we expect the distribution to be flat up to a maximum value given by the length of the quantum-mechanical spin ordering vector $\hat{\vec{m}}$, reflecting the simple model where the vector pointing in random directions on a sphere has a fixed length (Methods). 
The experimental data is in excellent agreement with \textit{ab initio} quantum Monte Carlo calculations of the Heisenberg model at the experimentally determined temperatures. 
These findings show that the lattice thermometer based on nearest-neighbour correlations employed here is correctly calibrated and very precise down to fractions of the tunnelling. 

While theoretical predictions at half-filling are available down to low temperature, this is not the case for doped systems \cite{Scalapino2006}. 
We can directly study the effect of doping on LRO in our experiment by reducing the density of our sample and measuring the spin structure factor. 
Within the region $\Omega$, we add a potential offset with the DMD for controlled hole doping.
The hole doping $\delta$ is deduced from the measured single particle density $n_s$ (Methods). 

\begin{figure}[t!]
\begin{center}
	\includegraphics{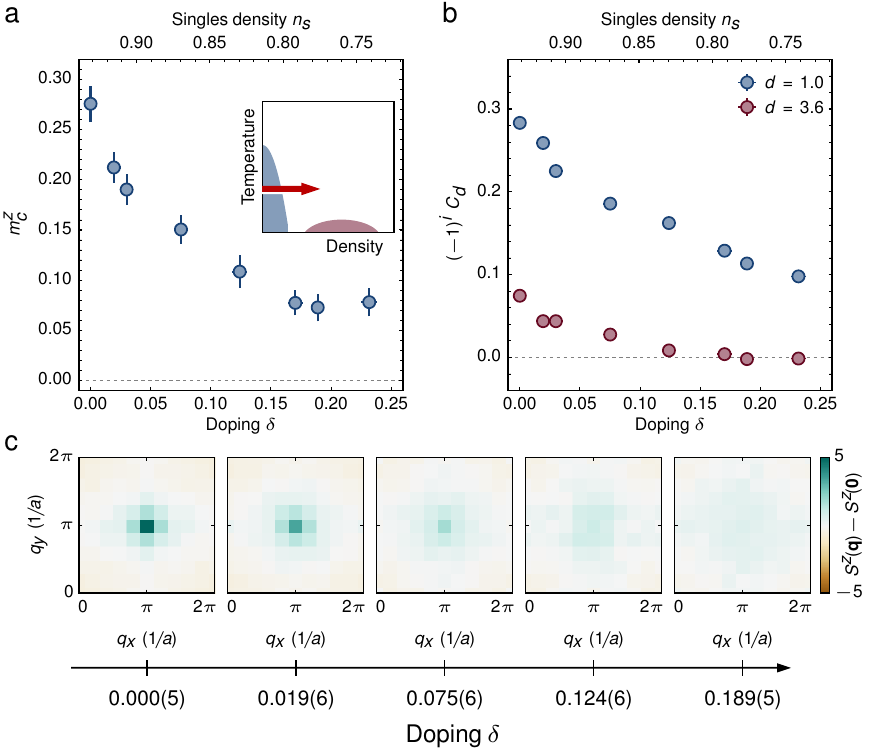}
\end{center}
	\caption{\textbf{Doping the antiferromagnet.}
		\textbf{a,} We move horizontally in the phase diagram by doping the system with holes (inset), where $0.0\lesssim \delta \lesssim 0.25$. The staggered magnetization $m^z_c$ settles at $\delta_c\approx 0.15$.
		\textbf{b,} The magnitude of the sign-corrected nearest-neighbour correlator decreases less rapidly with hole doping than correlators at larger distances. For large doping, only the nearest-neighbour correlator is appreciable, so this correlation is predominantly responsible for the non-zero staggered magnetization away from the antiferromagnetic phase. \textbf{c,} We show the spin structure factor, as in Fig.~2c, for each doping value. Error bars are computed as in (Methods).
		\label{fig:density}}
		\vspace{1cm}
\end{figure}

As shown in Fig.~\ref{fig:density}, we find that doping gradually suppresses $m^z_c$ and the weight of the antiferromagnetic ordering peak in $S^z(\vec{q})$.
Only at $\delta \gtrsim 0.15$ we find that $m^z_c$ settles to an approximately constant small value.
This offset originates from the strong short-range correlations still present at large dopings, see Fig.~\ref{fig:density}b.  
When excluding the contributions of $d<2$ from $m^z_c$, this offset disappears while the qualitative dependence $m^z_c(\delta)$ remains approximately the same, see Extended Data Fig.~5. 
This suggests that for the finite size $U/t=7.2$ Hubbard model studied here, antiferromagnetic LRO persists away from half-filling and disappears only above a critical hole doping $\delta_c\approx0.15$. 
This critical doping is large compared to values from empirically observed hole-doped cuprates of about $3\%$, and is actually closer to the optimal doping values for the high-temperature superconducting state \cite{Lee2006}.
But, unlike the cuprates, the system realized in our experiment is particle-hole symmetric and we expect the same critical value for particle doping.
For particle-doped cuprates, global antiferromagnetic order disappears around $14\%$, which is closer to our observed value. 
In this doping regime theoretical calculations of the Hubbard model suggest the existence of incommensurate antiferromagnetism, which would manifest as additional peaks in the spin structure factor at $\vec{q}\neq\vec{q_{\mathrm{AFM}}}$ \cite{Schulz1990}. 
For our parameters we observe no evidence for such order, possibly due to the finite system size. 

We have realized a long-range ordered quantum antiferromagnet governed by the 2D Hubbard Hamiltonian. 
Our architecture makes it possible to vary the doping and temperature, enabling us to explore the Hubbard phase diagram in theoretically challenging regimes. 
Attainable parameters are predicted to be sufficient to access the conjectured pseudogap \cite{Scalapino2006} and stripe ordered \cite{Lee2006} phases.
At lower temperatures $T/t \approx 0.05$ and dopings $\delta \approx 0.15$, theoretical work indicates a transition to a d-wave superconducting state \cite{Lee2006}. 
Such temperatures could be achieved through advanced entropy redistribution schemes \cite{Ho2009, Bernier2009, Ho2009a, Lubasch2011}.
The site-resolved control afforded by the microscope enables the measurement of entanglement and the engineering of dynamic structures to measure transport phenomena.
In particular, such structures allow out-of-equilibrium studies of time-resolved microscopic observables, such as the propagation of individually prepared holes, providing a deeper understanding of the many-body system.
Furthermore, entirely novel states of matter are within reach by augmenting the Hamiltonian with alternative lattice structures, artificial gauge fields, and dipolar long-range interactions.

\textbf{Acknowledgements} We thank Andreas Eberlein, Nikolai Prokov'ev, Subir Sachdev, Boris Svistunov, Wilhelm Zwerger, Ehud Altman, Mark Fischer and Martin Zwierlein and his research group for insightful discussions. 
We are grateful to Sebastian Blatt, R\'{e}mi Desbuquois, Susannah Dickerson, Adam Kaufman and Michael Messer for a critical reading of the manuscript. 
We are thankful to Richard Scalettar for illuminating discussions regarding the Hubbard quantum Monte Carlo theory. 
We thank Sebastian Blatt, Dylan Cotta, Simon F\"olling, Florian Huber, Widagdo Setiawan and Kate Wooley-Brown for early-stage contributions to the experiment. 
We acknowledge support from the Army Research Office Defense Advanced Research Projects Agency Optical Lattice Emulator Program, the Air Force Office of Scientific Research, the Multi University Research Initiative, the Office of Naval Research Defense University Research Instrumentation Program, and National Science Foundation. 
D. G. acknowledges support from the Harvard Quantum Optics Center and the Swiss National Science Foundation. 
A.M., C.S.C. and M.P. acknowledge support from the NSF. 
G. J. acknowledges support from NDSEG. 
R. S. is supported by the NSF through a grant for the Institute for Theoretical Atomic, Molecular and Optical Physics at Harvard University and the Smithsonian Astrophysical Observatory. 
F. G. gratefully acknowledges support from the Gordon and Betty Moore foundation. 
M. K.-N., F. G., and E. D. acknowledge support from NSF (Grant No. DMR-1308435), AFOSR Quantum Simulation MURI, ARO MURI on Atomtronics and AFOSR MURI Photonic Quantum Matter.

\textbf{Author contributions} All authors contributed extensively to the work presented here.

\textbf{Competing financial interests} The authors declare no competing financial interests.

%

\clearpage
\newgeometry{a4paper,
	left=0.75in,right=0.75in,top=0.75in,bottom=1.55in
}

\onecolumngrid

\begin{center} 
	\begin{Large} \textbf{Methods and Extended Data} \end{Large}
\end{center}
\medskip
\smallskip

\setcounter{section}{0}
\setcounter{subsection}{0}
\setcounter{figure}{0}
\setcounter{equation}{0}
\setcounter{NAT@ctr}{0}

\renewcommand{\figurename}[1]{FIG. }

\makeatletter
\makeatother
\renewcommand{\vec}[1]{{\boldsymbol{#1}}}
\renewcommand\Re{\operatorname{Re}}
\renewcommand\Im{\operatorname{Im}}

\subsection{State preparation.}

The low-temperature Fermi gas is prepared using a sequence similar to previous work, where a balanced mixture of the two lowest hyperfine states of $\Li$ with repulsive interactions is loaded into a 2D optical dipole trap \cite{Parsons2016}.
Prior to loading the lattice from the dipole trap we ramp to the final magnetic bias field used in the experiment, $576$\,G, corresponding to a scattering length of $210\,a_0$, and interaction energy $U/h= 6.50(3)$\,kHz. 
This includes a fast ramp over the narrow s-wave resonance at $543$\,G to avoid heating and loss. We then perform a final stage of evaporation immediately prior to loading the lattice, under a magnetic gradient. 
The magnetic gradient is then removed and we slowly load the atomic cloud within $40$~ms into a square optical lattice with tunnelling $t_x/h= 9.1(1)\times10^2$\,Hz and $t_y/h=9.0(1)\times10^2$\,Hz, with a lattice spacing of $a=569\,{\mathrm{nm}}$. 
The illumination of the DMD with blue-detuned light at wavelength $\lambda=650\,\mathrm{nm}$ is increased concurrently with the optical lattice. 
The populated layer lies at the focus of a high resolution microscope system, which allows site-resolved detection of the lattice occupation and, at the same time, enables us to augment the harmonic lattice trap by projecting an approximately ring shaped potential onto the atoms from a DMD in the image plane. 
The ring centre contains the correct potential to prepare the subsystem $\Omega$, while the rim of the ring is shaped to reduce the filling of the reservoir \cite{Liang2010}.

To vary the temperature of the sample, the atomic gas is held in the combined lattice and $650\,\mathrm{nm}$ blue-detuned light potential for a variable time. The final evaporation setpoint is chosen such that the final state is at half-filling (i.e. more atoms for higher temperatures). For the coldest temperature samples (zero hold time), approximately $400$ atoms remain in the trap at a temperature $T \approx 0.1\,T_\mathrm{F}$, where $T_\mathrm{F}$ is the Fermi temperature. To vary the density of the sample, the blue-detuned light field level is adjusted with the DMD, as described in the next section.

\subsection{Potential engineering.}

In addition to the underlying harmonic well generated by the optical lattice, an incoherent, blue detuned ring-shaped light field partitions the system into a sparsely filled reservoir and densely filled central region $\Omega$, see Fig.~1. 
The blue detuned light has a wavelength of $650$\,nm with a bandwidth of $10$\,nm. The light is generated by an SLED (Exalos, EXS210044) amplified by two tapered amplifiers (Eagleyard, EYP-TPA-0650-00250-2007-CMT02-0000) in series.
 
The digital micromirror device (DMD) is a flexible tool for projecting nearly arbitrary light fields on the atomic system \cite{Liang2010, Zupancic2016, Gaunt2012, Hueck2016}. The device is placed in the image plane of the system. This prevents direct access to the phase of the light field and makes it nearly impossible to produce single-site sized features. However, the image plane approach is technically robust and suitable for applying weak potentials on larger length scales. Despite the close detuning of the light field, the low intensity ensures that the single photon heating does not play a role on experimentally relevant timescales.

We use the TI Lightcrafter 6500 DMD evaluation board, featuring a $1920\times 1080$ pixel resolution with $7.56 \times 7.56$\,$\mu \mathrm{m}$ mirrors. During each experimental cycle, a pattern is loaded onto the DMD and the mirrors are fixed in place for the duration of the exposure. Since the DMD is a binary device, the desired pattern has to be converted into binary form using Floyd-Steinberg error diffusion.

The desired ring-shaped potential is a 2D piecewise defined function optimized experimentally, see Extended Data Fig.~1. The reservoir is created by a combination of a broad Gaussian peak that compensates the underlying harmonic confinement and a gradient term that compensates any residual potential gradients in the system. The central part that creates the $\Omega$ region is created by a depression in the light field and is curved to ensure the total potential seen by atoms within $\Omega$ is flat. Variants of the DMD-engineered potential are shown in Extended Data Fig.~1, simulated by applying a Fourier optics based transfer function to the input signal and accounting for the incident mode shape on the DMD. The flatness of the potential within $\Omega$ is critical to the realization of uniform antiferromagnetic LRO because the hole doping is seen to strongly affect magnetic correlations. We characterize the flatness of the potential by measuring the atomic density distribution within $\Omega$, shown in Extended Data Fig.~2. Azimuthal averaging reveals that the resulting density is flat to within $<4\%$ over approximately 80 sites. In order to explore controlled doping levels, the absolute offset of the central potential is tuned by varying the depth of the central depression in the DMD pattern, as seen in Extended Data Fig.~1. 

\subsection{Detection methods.}

Details regarding the spin-removal process, site-resolved imaging, and analysis of site-resolved images can be found in previous work \cite{Parsons2015, Parsons2016}. Spin removal begins by rapidly reducing the tunnelling through increasing the lattice depth to $50\,E_r$. To maximize the singles density for the temperature dataset, the lattice depth ramp occurs over $5\,\mathrm{ms}$, whereas for the density dataset, the ramp occurs over $0.8\,\mathrm{ms}$ to reduce density fluctuations for a given density setpoint. 

For each temperature value in the temperature dataset and for each doping value in the doping dataset, we take $70$ images with no spin removed, $70$ with one spin removed, and $70$ with the other spin removed. 

\subsection{Data analysis.}

All quantities presented in this letter are extracted from a circular section of $10$ lattice sites in diameter, denoted $\Omega$.
This disk is centred at the centre of mass of sites exceeding $80\%$ filling, as measured from the images with no spin removal.
Sites are included in $\Omega$ if the centre of the site is within the bounds of the circle.
Since the centre of mass is a non-integer coordinate, our system sizes vary between $75$ and $81$ sites depending on the exact location of the centre within a lattice site.
We find that in all cases, the filling is constant across the disk.
The entire region $\Omega$ is then used for all calculations of spin correlations, spin structure factor, and staggered magnetization. 
We determine the temperature of each sample by comparing the largest measured nearest-neighbour correlator in $\Omega$ to quantum Monte Carlo predictions at half-filling \cite{Paiva2010}.

We calculate the average site-resolved spin correlator for a given displacement, $C_{\vec{d}}$, by averaging the correlator over all pairs of sites within $\Omega$ with the given displacement.
We extract this quantity using the alternative method described in previous work \cite{Parsons2016}, due to it having lower uncertainty for small values of $C_{\vec{d}}$.
Displacement vectors that have less than 100 total pairs of sites across all images are discarded.
To calculate the spin correlation function versus distance, the displacements are split into bins of 0.3 sites, and averaged across.
Because of the non-integer centre of $\Omega$, the largest distances may appear infrequently, so we discard bins for distances of larger than 5 sites if there are less than 5 contributions in that bin.
Errors on the spin correlator are calculated as in previous work \cite{Parsons2016}, and the errors on the spin correlation function are propagated using conventional techniques assuming the spin correlation errors are distributed normally. 

For the structure factor we first inscribe the circular section $\Omega$ into a minimal-size square (see Data Analysis). 
In the associated occupation matrix atoms are denoted as 1, unoccupied sites denoted as -1, and sites introduced from placing the circular section in a square array denoted as 0.
We compute the magnitude squared of the Fourier transform of this array and average across all experimental realizations,
\begin{equation}
\begin{aligned}
\langle {\left|\mathcal{F}(\vec{q})\right|}^2 \rangle &= \left\langle \sum_{\vec{r}, \vec{s}}\mathrm{e}^{i \vec{q}\cdot(\vec{r}-\vec{s})}(\hat{S}^z_{\vec{r}}+(\hat{m}^z_{0,\vec{r}})^2-1)(\hat{S}^z_{\vec{s}}+(\hat{m}^z_{0,\vec{s}})^2-1) \right\rangle\\
&= N S^z(\vec{q}) + \bigg|\sum_{\vec{r}}\mathrm{e}^{i \vec{q}\cdot\vec{r} } \left\langle(\hat{m}^z_{0,\vec{r}})^2-1\right\rangle\bigg|^2\\
\Longrightarrow S^z(\vec{q}) &= \frac1{N} \left(\langle {\left|\mathcal{F}(\vec{k})\right|}^2 \rangle - \bigg|\sum_{\vec{r}}\mathrm{e}^{i \vec{q}\cdot\vec{r} } \left\langle1-(\hat{m}^z_{0,\vec{r}})^2\right\rangle\bigg|^2\right),
\label{methods:ssf}
\end{aligned}
\end{equation}
where $N$ is the number of sites, $S^z(\vec{q})$ is the spin structure factor, $\hat{S}^z_{\vec{r}}$ is the spin operator at site $\vec{r}$, $(\hat{m}^z_{0, \vec{r}})^2$ is the local moment operator, and due to spin balance $\langle \hat{S}^z_{\vec{r}} ((\hat{m}^z_{0,\vec{s}})^2-1) \rangle = 0$.
Errors are calculated by assuming the distribution of $S^z(\vec{q})$ is normal.
Though there are finite bounds on the possible values of the spin structure factor, the histograms in Fig.~3 demonstrate that the assumption of normality is reasonable. 

In the thermodynamic limit, the RMS staggered magnetization is
\begin{equation}
m^z = \sqrt{\frac{1}{N} S(\vec{q}=(\pi/a, \pi/a))}.
\end{equation}
For finite systems this expression is non-zero even in the paramagnetic limit.
We remove the finite offset and normalize such that the minimum value is $0$ in the absence of spin correlations and the maximum value is unity:
\begin{equation}
m^z_c=\sqrt{|S^z(\vec{q}_{\mathrm{AFM}})-S^z(\vec{0})|} \sqrt{N/(N^2-N)}.
\end{equation}
Errors are propagated from the computed spin structure factor using standard techniques.
We verify that the value resulting from the Fourier-based computation agrees with the value obtained by summing individual site-resolved spin correlations, which explicitly accounts for doublons and holes in the sample, see Extended Data Fig.~3 (due to spin balance $\hat{S}_{\vec{r}}^z=0$).

\subsection{System calibrations.}

The system is calibrated via parametric lattice modulation spectroscopy, where the lattice is modulated at varied frequencies. When the modulation frequency matches an inter-band transition, significant loss of atoms is seen. The inter-band transition frequencies are fitted to a Mathieu equation model, and a conversion between lattice laser power and lattice depth is extracted, which sets the tunnelling parameter $t$. The interaction strength $U$ is estimated from the scattering length at the applied magnetic field and the lattice depths along the $x, y$ and $z$ directions. 

\subsection{Theoretical methods.}

In the region $\Omega$, our system is well-described by the single-band, homogeneous, two-component Hubbard model on a square lattice \cite{Hubbard1963}:
\begin{equation}
\hat{H} = -t \sum_{\sigma, \left\langle i, j \right\rangle \in \Omega}{(\hat{c}^{\dagger}_{i,\sigma}\hat{c}_{j,\sigma} + h.c.)} + U\sum_{i \in \Omega}{\hat{n}_{i,\downarrow}\hat{n}_{i,\uparrow}},
\end{equation}
where $\left\langle i, j \right\rangle$ are nearest-neighbours and $\hat{c}_{i,\sigma}$ and $\hat{n}_{i,\sigma} = \hat{c}^{\dagger}_{i,\sigma}\hat{c}_{i,\sigma}$ are the annihilation and number operator of spin $\sigma \in \{\uparrow, \downarrow\}$ on site $i$, respectively. The on-site spin operators are defined $\hat{S}^\alpha_i = \frac12\hat{c}^{\dagger}_{i,\beta}\sigma^{\alpha}_{\beta\gamma}\hat{c}_{i,\gamma}$ where $\sigma^{\alpha}$ are the Pauli spin matrices.

Close to half-filling and at temperatures significantly below the interaction energy, the Hubbard model exhibits a linear relation between a change of the singles occupation $ \delta_{\text{singles}}$ and doping of the total density $\delta$. Using numerical data obtained from the dynamical cluster approximation \cite{Leblanc2013} method at $U/t = 7.2$ and $T/t=0.25$, we obtain a fitted slope of $\delta= 1.22(1) \times \delta_{\text{singles}}$. The relation we obtain is consistent with results obtained from a resummed numerical linked-cluster expansion \cite{Paiva2010} (NLCE) method for dopings less than $6\%$, above which the resummed NLCE data becomes unstable. We use this result to calculate the hole doping given the singles doping in Fig.~4.

The quantum non-linear $\sigma$ model mentioned in the main text was originally introduced to describe the low-temperature Heisenberg model \cite{Chakravarty1989,Chubukov1994}. Finite size effects are expected to increase the spin stiffness $\rho_s$  \cite{Caffarel1994}.

For the theory prediction in Fig.~2d we perform determinant quantum Monte Carlo on the $10\times10$ periodic-boundary 2D Hubbard Model at $U/t = 7.2$ using the QUEST package \cite{Chang2015, Gogolenko2014, Jiang2016}. At low temperatures, the system does not isotropically sample different orientations of the staggered magnetization within the $10^4$ measurement sweeps, so the reported magnetization is averaged over 14 independent runs and averaged over all three axes.

\subsection{Theoretical methods: Histograms.}

Full counting statistics (FCS) represent a powerful tool to characterize quantum states and phenomena in a variety of systems \cite{Cherng2007, Braungardt2008, Braungardt2011, Lamacraft2007}. For example, it has been used to observe the quantization of electrical charge in shot noise measurements \cite{Blanter2000}, the observation of fractional charges in fractional quantum Hall systems \cite{Picciotto1997, Saminadayar1997, Goldman1995}, and was used to characterize prethermalization in an ultracold atomic setup \cite{Gritsev2006, Polkovnikov2006, Hofferberth2008}. Here we determine the FCS of the staggered magnetization operator $\hat{m}^z = \frac{1}{N} \sum_i (-1)^i \frac{1}{S} S_i^z$ from an \textit{ab initio} quantum Monte Carlo simulation of the antiferromagnetic Heisenberg model 
\begin{equation}
\hat{H} = J \sum_{\langle i, j \rangle}\hat{\vec{S}}_i\cdot\hat{\vec{S}}_j.
\end{equation}
To this end we implement a stochastic series expansion quantum Monte Carlo calculation with operator loop updates \cite{Sandvik1999}. We simulate a $16\times16$ system with periodic boundary conditions, and calculate the FCS of $\hat{m}^z$ in a smaller $9\times9$ region, which is of similar size as the measurement region used in the experiment. In our model the spins outside of the measurement area serve as an effective thermal bath, mimicking the experimental setup. The presence of the bath introduces additional fluctuations compared to a system of equal size but with periodic boundary conditions, leading to a further suppression of large values of the staggered magnetization at the lowest temperatures in Fig~\ref{fig:histogram}.

As shown in Figure \ref{fig:histogram}, the distribution of measured values of $\hat{m}^z$ shows qualitatively different characteristics at high and low temperatures, imposed by the infinite and zero temperature limits. At infinite temperature, the spins take on uncorrelated random values. In a system of $N$ measured sites, the FCS has thus the same distribution as $N$ independent coin flips, which approaches a binomial distribution of width $1/\sqrt{N}$ for large system sizes.

\subsection{Alternative basis measurements.}
All measurements in this paper were carried out in the $\hat{S}_z$ basis. SU(2) symmetry predicts that correlations would exist along every possible measurement axis. To verify this, we added a $\pi/2$ pulse prior to the selective spin removal pulse and measured correlations as before. Extended Data Fig.~4 shows that the correlations at all length scales are insensitive to the applied RF field, regardless of whether no pulse, a $\pi/2$ pulse, or a $\pi$ pulse is performed. This outcome is consistent with but not sufficient for a claim that the underlying state is SU(2) symmetric.	  
\FloatBarrier
\newpage

\begin{figure}
	\centering	\includegraphics[width=\linewidth]{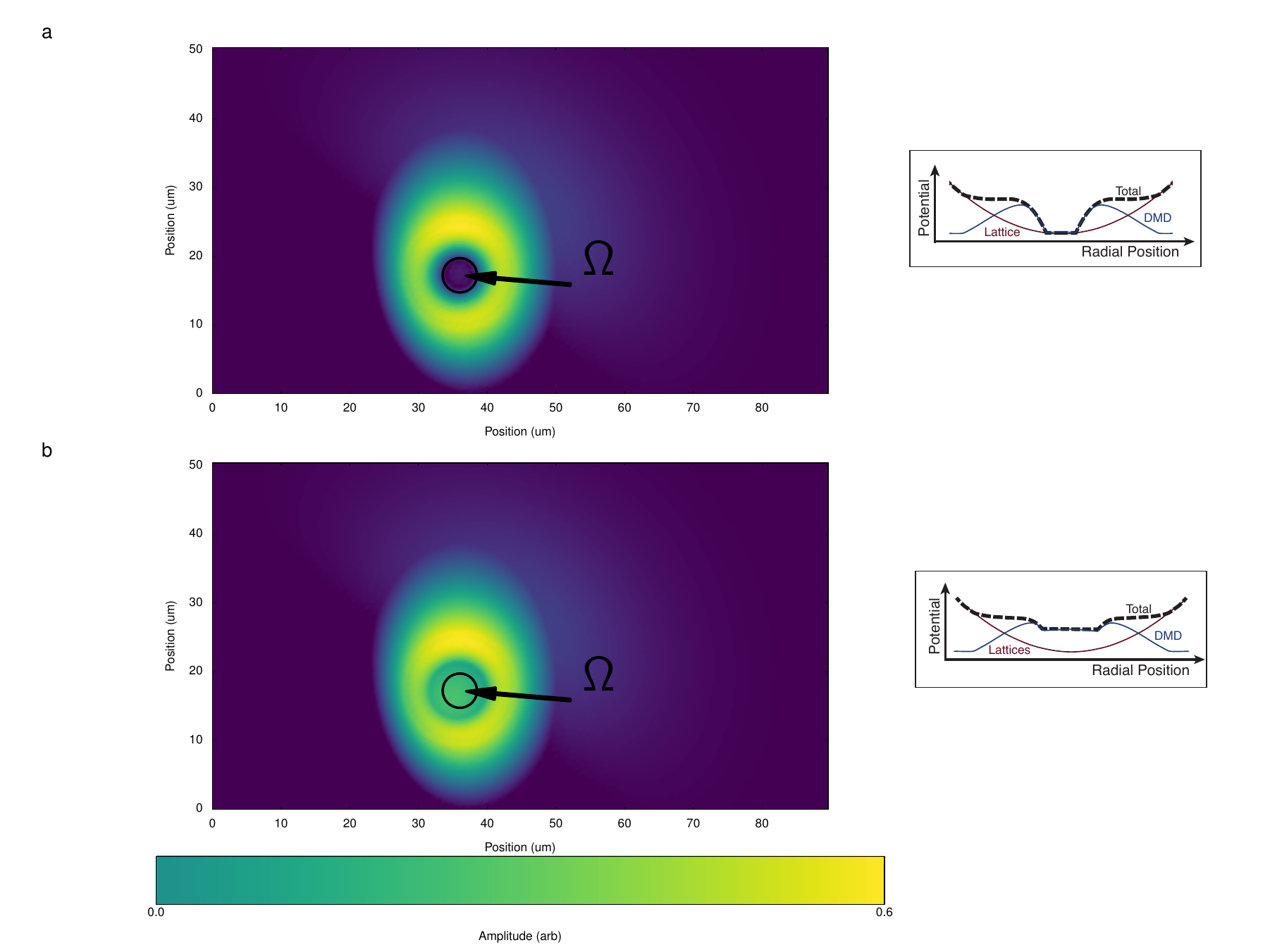}
	\label{fig:profile}
	\caption{\textbf{Amplitude of light fields applied to atoms} \textbf{a,} The computed light field generated by the DMD, applied to the atoms for half-filled samples. A gradient compensates residual gradients in the lattice. The rim of the donut provides sharp walls for the inner subsystem. A small peak in the center flattens the potential when combined with the optical lattice. The cartoon shows a schematic view of a radial cut of the potential, including the contribution of the lattice. \textbf{b,} The amplitude of the light field with an offset in the center of the trap, used to dope the system with a finite population of holes.}
\end{figure}

\begin{figure}
	\centering	\includegraphics[width=\linewidth]{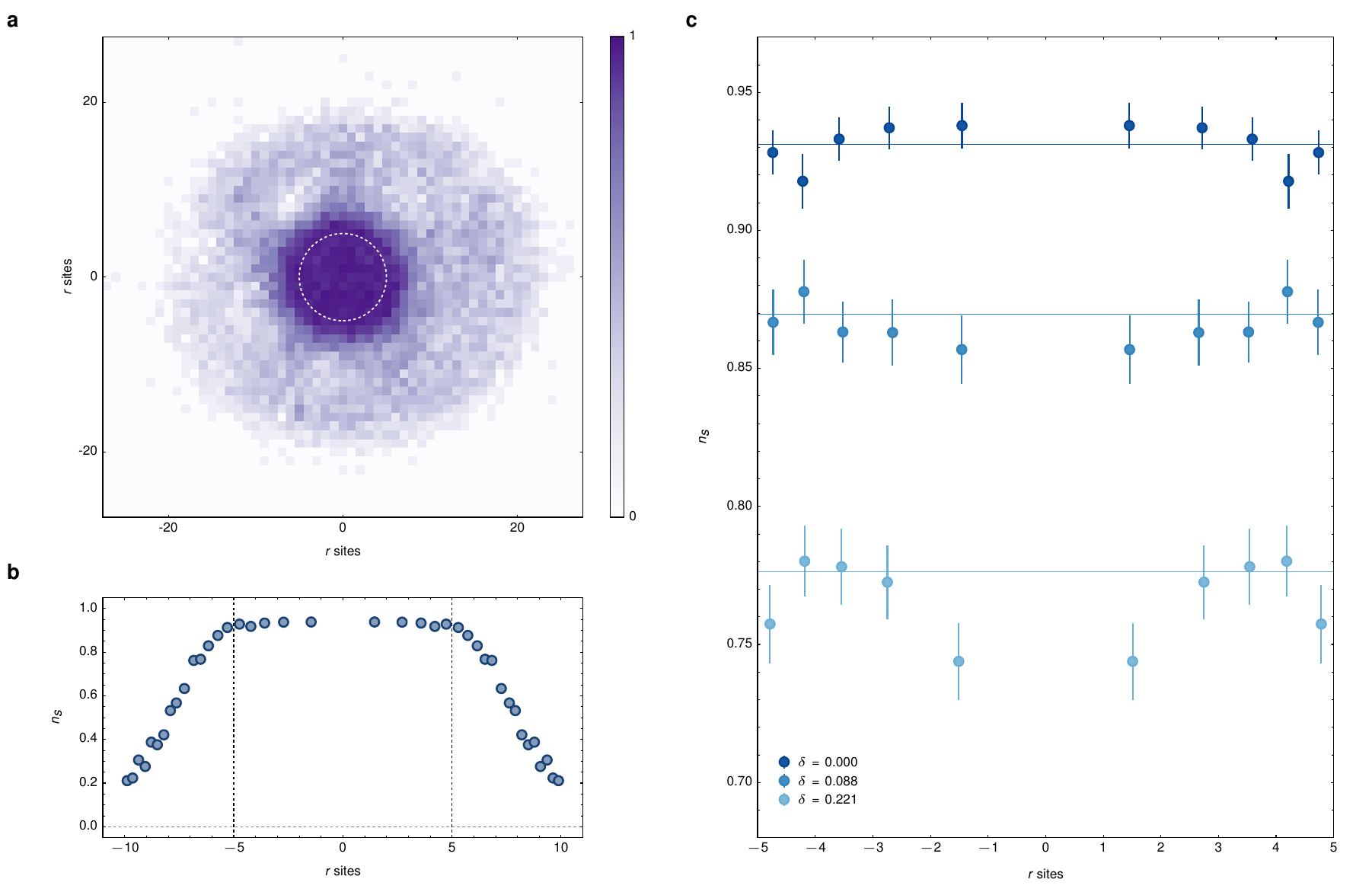}
	\label{fig:atom_density}
	\caption{\textbf{Average density profile in the system.} \textbf{a,} The average singles density map for a sample at half-filling shows a central region of uniform density, surrounded by a donut-shaped ring of low density. The dotted white circle indicates our system size, excluding edge effects. \textbf{b,} The azimuthal average of the singles density shown in \textbf{a}, for the system as well as the inner edge of the donut where the density drops off to the reservoir density. The vertical dotted lines denote the boundary of the system. \textbf{c,} Azimuthal average of singles densities for three values of the hole doping used in the experiment, indicating uniformity of atom number across our system to within 4\%. The horizontal lines are at the system-wide average densities.}
\end{figure}

\newpage

\begin{figure}
	\centering
	\includegraphics{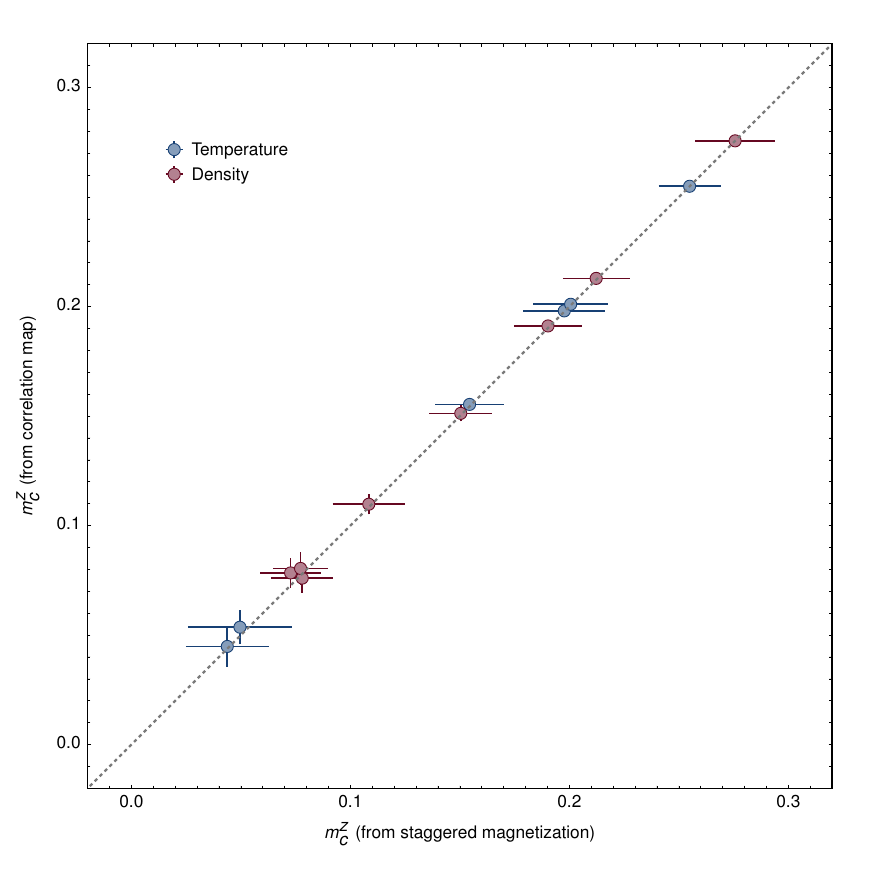}
	\caption{\textbf{Comparison of staggered magnetizations obtained directly through single-spin images and from spin correlations.} We calculate the staggered magnetization from images with one spin state removed (main text). The staggered magnetization can also be calculated from the spin correlator (methods), where the two are exactly equal in the limit of no noise and exactly one particle per site. Plotting these two quantities against each other, we find very good agreement with the line $y=x$ (dotted line), indicating that any error due to deviation from one particle per site is small.}
\end{figure}

\newpage
\begin{figure}
	\centering	\includegraphics{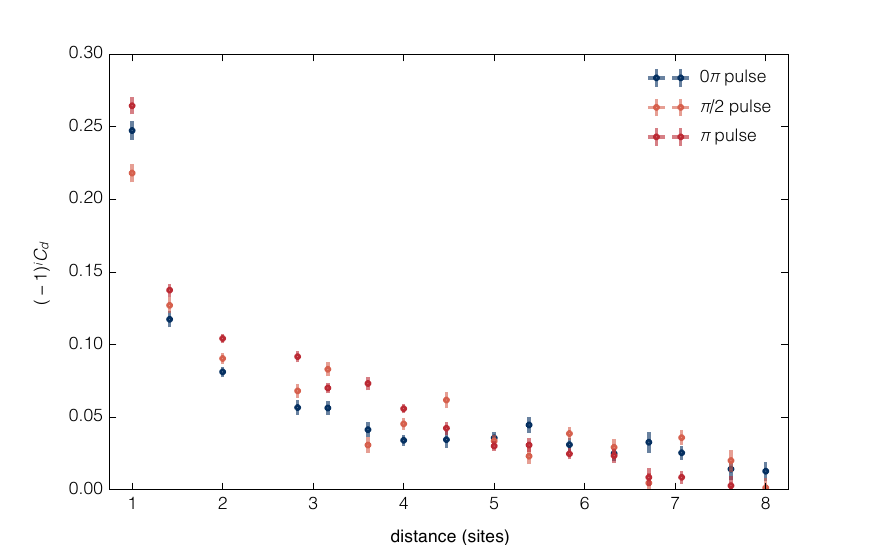}
	\label{fig:microwaves}
	\caption{\textbf{Alternative basis measurement.} We optionally apply a $\pi/2$ or $\pi$ RF pulse prior to the spin removal pulse and correlation measurement. The sign-corrected spin correlation functions show an insensitivity to the presence and duration of this RF pulse, consistent with an SU(2) symmetry of the state.}
\end{figure}

\newpage

\begin{figure}
	\centering
	\includegraphics{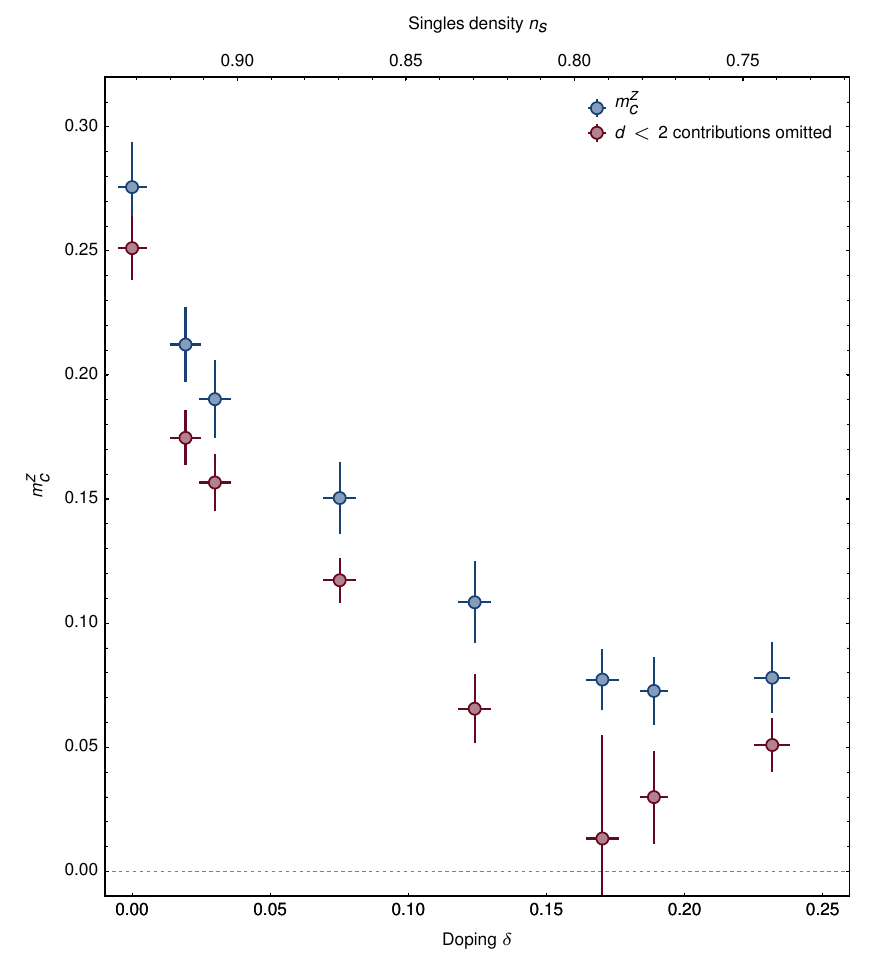}
	\caption{\textbf{Staggered magnetization obtained from spin correlations, with and without nearest-neighbor contribution included.} To investigate the contributions to the staggered magnetization at high dopings, we consider the staggered magnetization calculated from the spin correlator. We omit the longest-range correlations, which have the greatest level of noise due to the low number of pairs of sites extending across the cloud, as well as nearest-neighbor correlations, which is essentially the only non-zero correlator outside of the AFM phase. In the high-doping regime, we see that the greatest contribution to the staggered magnetization is the nearest-neighbor correlation, followed by the noisy longest-range correlations. The value of the staggered magnetization without either of these contributions is plotted in red.}
\end{figure}

\newpage

\end{document}